\documentclass[aps,preprint,amsmath,amssymb,showpacs,prb]{revtex4-1}
\usepackage[utf8x]{inputenc}
\usepackage{graphicx}
\usepackage{epsf}
\usepackage{natbib}

\usepackage{graphicx}
\usepackage{units}
\usepackage[mediumspace,mediumqspace,squaren]{SIunits}

\newcommand{\reffig}[1]{Fig.\ref{#1}}

\newcommand{\PP}[1]{probe particle }

\begin{document}

\title{The mechanism of high-resolution STM/AFM imaging with functionalized tips}

\author{Prokop Hapala}
\email[corresponding author: ]{hapala@fzu.cz}
\affiliation{Institute of Physics, Academy of Sciences of the Czech Republic, v.v.i.,  Cukrovarnick\' a 10, 162 00 Prague, Czech Republic}
\author{Georgy Kichin}
\affiliation{Peter Gr{\"u}nberg Institut (PGI-3),Forschungszentrum J{\"u}lich, 52425 J{\"u}lich, Germany}
\affiliation{ J{\"u}lich Aachen Research Alliance (JARA)--Fundamentals of Future Information Technology, 52425 J{\"u}lich, Germany}
\author{Christian Wagner}
\affiliation{Peter Gr{\"u}nberg Institut (PGI-3),Forschungszentrum J{\"u}lich, 52425 J{\"u}lich, Germany}
\affiliation{ J{\"u}lich Aachen Research Alliance (JARA)--Fundamentals of Future Information Technology, 52425 J{\"u}lich, Germany}
\author{F. Stefan Tautz}
\affiliation{Peter Gr{\"u}nberg Institut (PGI-3),Forschungszentrum J{\"u}lich, 52425 J{\"u}lich, Germany}
\affiliation{ J{\"u}lich Aachen Research Alliance (JARA)--Fundamentals of Future Information Technology, 52425 J{\"u}lich, Germany}
\author{Ruslan Temirov}
\email{r.temirov@fz-juelich.de}
\affiliation{Peter Gr{\"u}nberg Institut (PGI-3),Forschungszentrum J{\"u}lich, 52425 J{\"u}lich, Germany}
\affiliation{ J{\"u}lich Aachen Research Alliance (JARA)--Fundamentals of Future Information Technology, 52425 J{\"u}lich, Germany}
\author{Pavel Jel\'{i}nek}
\affiliation{Institute of Physics, Academy of Sciences of the Czech Republic, v.v.i.,  Cukrovarnick\' a 10, 162 00 Prague, Czech Republic}
\affiliation{Graduate School of Engineering, Osaka University 2-1, Yamada-Oka, Suita, Osaka 565-0871, Japan}


\begin{abstract}
High resolution Atomic Force Microscopy (AFM) and Scanning Tunnelling Microscopy (STM)  imaging with functionalized tips is well established, but a detailed understanding of the imaging mechanism is still missing. We present a numerical STM/AFM model, which takes into account the relaxation of the probe due to the tip-sample interaction. We demonstrate that the model is able to reproduce very well not only the experimental intra- and intermolecular contrasts, but also their evolution upon tip approach. At close distances, the simulations unveil a significant probe particle relaxation towards local minima of the interaction potential. This effect is responsible for the sharp sub-molecular resolution observed in AFM/STM experiments. In addition, we demonstrate that sharp apparent intermolecular bonds should not be interpreted as true hydrogen bonds, in the sense of representing areas of increased electron density. Instead they represent the ridge between two minima of the potential energy landscape due to neighbouring atoms.
\end{abstract}

\maketitle

\section{Introduction}

Scanning Tunnelling (STM) and Atomic Force Microscopy (AFM) methods are key tools of nanoscience. One of their most remarkable achievements is the unprecedented sub-molecular resolution of both atomic and electronic structures of single molecules on surfaces. First real space images of molecular orbitals were obtained with STM \cite{Repp_PRL05}. Then it was found that functionalizing the tip with a single carbon monoxide (CO) molecule enhances the resolution of molecular orbital STM images \cite{Bartels_PRL98, Gross_PRL2011}. Later it has been discovered that STM tip functionalization with H$_2$, D$_2$ (so-called scanning tunnelling hydrogen microscopy (STHM) \cite{Temirov2008, Weiss2010a}) and a variety of other atomic and molecular particles (Xe, CH$_4$, CO) allows the STM to resolve the \textit{atomic} structures of large organic adsorbates in a direct imaging experiment \cite{Temirov2008, Weiss2010a, Kichin2011, Kichin2013}. At the same time, the development of the qPlus AFM technique \cite{Giessibl2003} 
has resulted in the successful resolution of internal molecular structures by AFM \cite{Gross_Science09, Gross2012}.

On the basis of a density functional theory (DFT) analysis the high resolution of molecular structures in AFM has been attributed to Pauli repulsion  \cite{Gross_Science09,Moll_NJP10}. Following this result it has been proposed that the contrast delivered by functionalized STM tips is related to the same force \cite{Weiss2010a}.  One peculiar feature of the high-resolution STM/AFM images obtained with functionalized tips, namely the striking imaging contrast obtained in areas between molecules \cite{Weiss2010, Kichin2011, Kichin2013, Zhang2013}, has, however, not yet been clearly explained. In particular, the sharp ridges observed in the high resolution images do not necessarily represent a~true bond. For example, Pavlicek et al \cite{Pavlicek2013} observed in high resolution images of DBTH molecules an~apparent bond ridge between sulphur atoms where there is no chemical bond. 

Very recently it has been argued that sharp contrast features between the molecules may be related to the imaging of hydrogen bonds \cite{Zhang2013},  as a consequence of an enhanced electron density between oxygen and hydrogen atoms of neighbouring molecules. But it is not clear why in experiment this contrast appears so sharp, especially since according to DFT simulations the electron density variation is expected to be exceedingly small \cite{Zhang2013}. The sharp lines visible in these experiments therefore cannot be automatically ascribed to (hydrogen) bonds. This calls for a deeper understanding of the origin of the high resolution contrast in general. 

In this communication we propose a simple mechanical model of the functionalized STM/AFM junction that for the first time clarifies all of the main features of STM and AFM images measured with functionalized tips.  First, it explains the appearance of characteristic sharp features in STM and AFM images measured at close tip-sample distances. Secondly, it establishes the relationship between the observed AFM and STM image contrasts obtained with functionalized tips. Thirdly, it reveals the nature of the STM and AFM contrasts in the intermolecular regions and allows a critical discussion of the appearance of so-called hydrogen bonds in the images. Finally, the method allows us to simulate AFM/STM images of complete molecular layers at different tip-sample distances at small computational cost. To underpin the predictive power of our model, we compare AFM/STM images obtained from our numerical model to selected experimental cases.

\begin{figure}
\centering
\includegraphics[width=8.5cm]{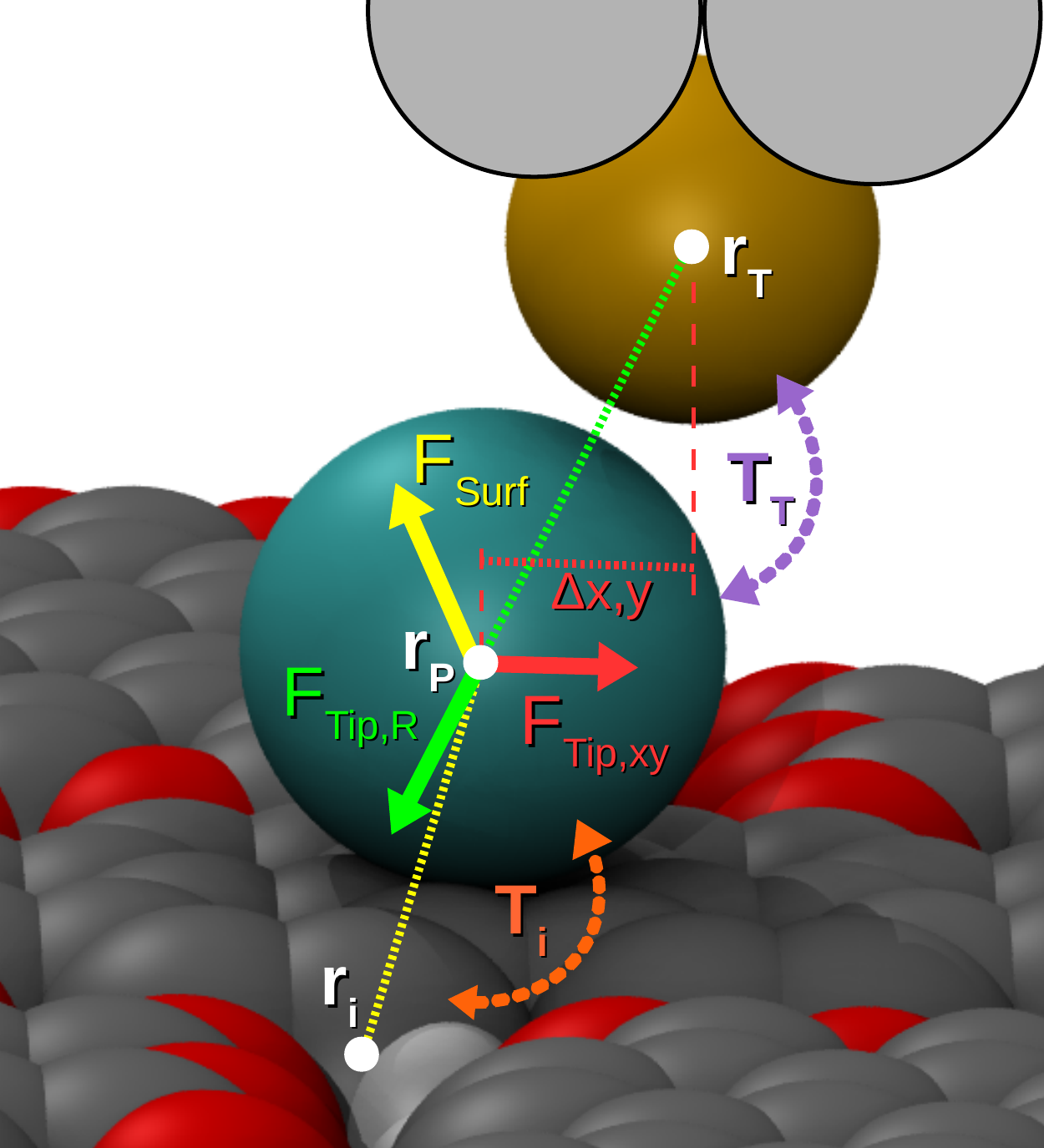}
\caption{\textbf{AFM/STM model}. Schematic view of the mechanical model of a functionalized tip as employed in this work. The last metal atom of the tip ({\it tip base}) is shown in sand colour, the \PP\ in cyan and the molecular layer ({\it sample}, in the example a herringbone PTCDA layer) in grey (carbon atoms) and red (oxygen atoms).  The forces acting on the probe particle are shown in colour: Radial tip force $F_{\mathrm{Tip,R}}$ (green); lateral tip force $F_{\mathrm{Tip,xy}}$ (red); force $F_{\mathrm{Surf}}$ exerted by atoms of the sample (yellow). Force-determining geometric parameters are shown in the same colour as the corresponding forces. The two distinct hopping processes in our STM model are denoted by violet (probe particle-tip, T$_T$) and orange (probe particle-sample, T$_i$) colour.}
\label{fig-01}
\end{figure}



\section{Methods}
\subsection{Equipment and sample preparation:}
The experiments were performed using a combined NC-AFM/STM from Createc. The base pressure at the working temperature of 5 K was better than 10$^{-11}$ mbar. All samples were prepared using standard techniques of surface preparation in ultra-high vacuum. Sub-monolayer coverages of PTCDA were deposited onto freshly prepared surfaces of Au(111) and Ag(111) at room temperature using a home-built Knudsen-cell. Immediately after the deposition the sample was transferred to the cold (5 K) STM.  Individual xenon atoms and carbon monoxide molecules were deposited onto the sample at 5 K by closing the ion getter pump, opening the shutter in the cryoshields, and flooding the STM chamber with the clean gas for 10 minutes at a pressure of about $5 \times 10^{-9}$ mbar. The tip decoration was effected according to the procedures described in reference \cite{Kichin2011}.

\subsection{Preparation molecular geometry for simulation:} 
The molecular geometry of PTCDA monolayers on the respective metallic substrates was determined from  supercell parameters obtained from reference\cite{Kilian2006} in case of Au(111) and from reference \cite{Mercurio2013} for Ag(110). Atomic coordinates for the 8-hydroxyquinoline tetramer were taken from reference \cite{Zhang2013}. The molecular geometries were then relaxed in the relevant supercell using the local orbital DFT code FIREBALL \cite{Jelinek2005PRB, Review-Fireball2011} within the local-density approximation (LDA) for the exchange-correlation functional. During the relaxation the molecular geometry was free to move in the $x,y$ plane, while the $z$ coordinates of all atoms were set to $z=0$ and fixed. Convergence was achieved once a residual total energy of 0.0001 eV and a maximal force of 0.05 eV/ \AA\ were reached. We took atomic structure of NTCDI molecules deposited on Ag:Si(111)-($\sqrt3 \times \sqrt3$) from DFT calculations published in paper \cite{Sweetman2014}  (see Figure 3.C in the 
reference ).

\subsection{Mechanical model:} 

The Lennard-Jones parameters used in the mechanical model of \PP\ relaxation for H, C, O, Xe atoms were taken from the OPLS force field \cite{Jorgensen1988} (for more details see Supplementary Table 1 and Supplemenatry Methods). The robustness of the simulations with respect to the precise values of these parameters is demonstrated in Supplementary Figures 1, 2, and 3. Regarding the procedure in which the simulated AFM images were generated, refer to Supplementary Methods. 

The main ingredient of our model is the geometric distortion of the `soft' apex of a functionalized tip due to the interaction with the surface \cite{Gross2012}.  We model this soft apex as the outermost atom of the metal tip ({\it tip base}) and the {\it probe particle} that decorates it (see \reffig{fig-01}).  

To account for the interaction between the functionalized tip and a molecular layer (`sample') on the surface we construct a force-field model of the junction using empirical potentials. In particular, we use a pairwise Lennard-Jones (L-J) potential to describe the weak interaction $F_{\mathrm{Surf}}$ between the \PP\ of the functionalized tip and the sample (see \reffig{fig-01}). $F_{\mathrm{Surf}}$ is calculated as a~sum of all pairwise L-J forces acting between the \PP\ and the atoms constituting the molecular layer. Besides $F_{\mathrm{Surf}}$ the \PP\ experiences two additional forces: (i) A radial L-J force $F_{\mathrm{Tip,R}}$ between the \PP\ and the tip base which keeps the \PP\ attached to the tip base at a particular distance and (ii) an additional lateral harmonic force $F_{\mathrm{Tip,xy}}$ that stems from the cylindrically symmetric attractive potential of the tip base. 

In this work we employ two different sets of L-J parameters (binding energy $\epsilon_\alpha$ and equilibrium distance $r_\alpha$) of the $F_{\mathrm{Tip,R}}$ interaction to mimic Xe- and CO-decorated tips (cf.~Supplementary Table 1), while the lateral stiffness $k_{xy} = \unit{0.5}{\newton\per\metre}$ is kept constant for all types of probe particles. Interestingly, we find that the results depend only weakly on variations of the binding energy parameter $\epsilon_\alpha$ (cf.~Supplementary Figure 1). This observation agrees well with the fact that the high-resolution images of a particular molecule obtained with different functionalized tips look qualitatively similar \cite{ GrossAPL2013}. It turns out that the variation of the image contrast between different probe particles is mainly related to their different van der Waals radii, in our model defined by $r_\alpha$ (cf.~Supplementary Figure 2). For further information on the employed L-J parameters see Supplementary Methods.


We use our model to simulate the images of the~well-known herringbone monolayer of 3,4,9,10-perylene tetracarboxylic dianhydride (PTCDA). PTCDA layers, as well as single PTCDA molecules, have been extensively imaged with functionalized STM/AFM tips \cite{Temirov2008,Weiss2010a,Weiss2010,MohnPRL2010,Kichin2011,Kichin2013}. Therefore, a wealth of experimental data is readily available for direct comparison to the results of our simulations. The input atomic structure of the molecular layer was taken from the data published for PTCDA/Au(111) \cite{Kilian2006}. The structure was further optimized with DFT. The simulated data were acquired by scanning the model tip laterally over the surface with a step of $\Delta x, \Delta y = \unit{0.1}{\angstrom}$. 

At each lateral position, the tip was placed at an~initial set point $z_0=\unit{12}{\angstrom}$ above the molecular layer.  Subsequently we approached the tip vertically towards the sample in steps of $\Delta z = \unit{0.1}{\angstrom}$ until $z=\unit{6}{\angstrom}$. At each step of the vertical approach the \PP\ 's position was allowed to relax until the net force $F_{\mathrm{Surf}}+F_{\mathrm{Tip,R}}+F_{\mathrm{Tip,xy}}$ acting on the \PP\ (see \reffig{fig-01}) became smaller than $\unit{10^{-6}}{\electronvolt\per\angstrom}$. At the same time the degrees of freedom of the molecular layer and the tip base were kept fixed. Once the structural relaxation was completed, the vertical force $F_z$ was calculated from a~projection of $F_{\mathrm{Surf}}$ onto the $z$ axis. Finally, the $F_z(z)$ curves were converted to frequency shift $\Delta f(z)$ using the inverse Sader formula \cite{Sader2004}. 

\begin{figure}
\centering
\includegraphics[width=6.5cm]{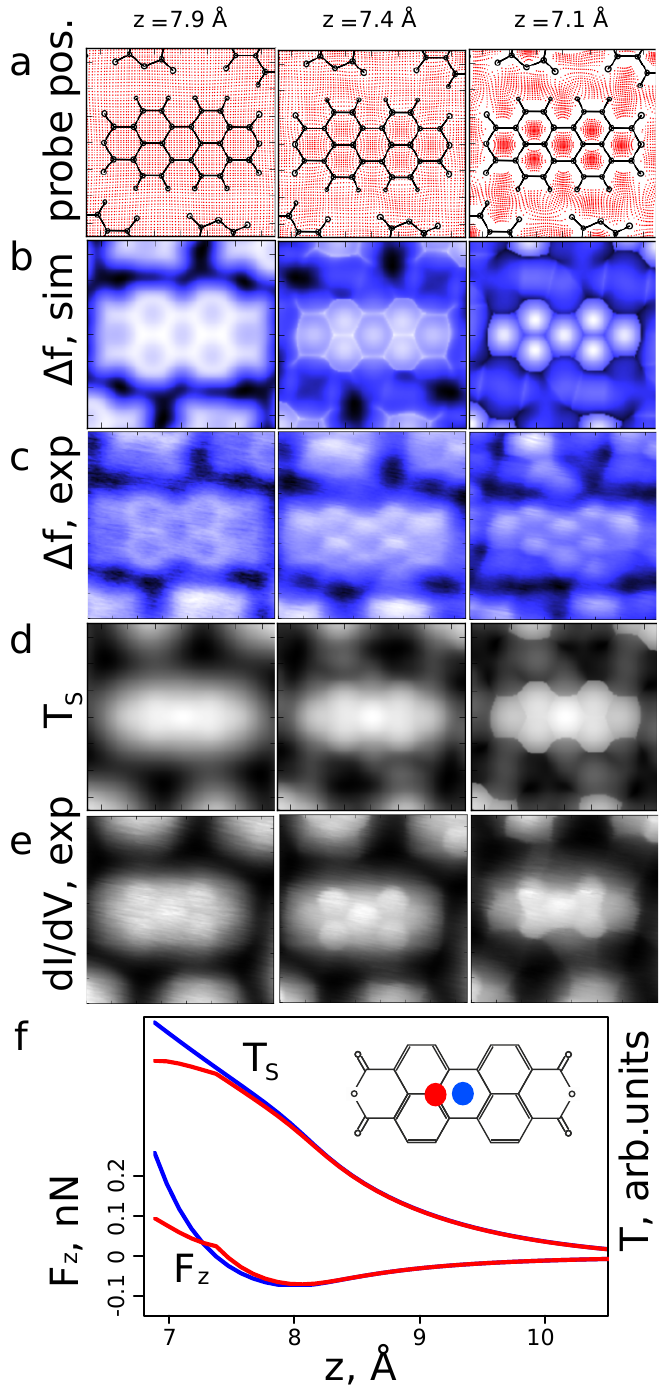}
\caption{\textbf{ Simulated and experimental high resolution AFM and STM images.}  Experimental images have been recorded on PTCDA/Au(111) with a CO probe particle. Simulated images have been obtained with the following L-J parameters mimicking a CO probe particle. (a) Map of simulated \PP\ positions after relaxation. (b) Simulated AFM image (the frequency shift ${\Delta}f$ is displayed). (c) Experimental AFM images (frequency shift). (d) Simulated STM images (maps of the $T_S$ tunnelling process). (e) Experimental STM images (differential conductance). (f) Vertical force $F_z$ (left axis) and tunnelling probability $T_S$ (right axis, arbitrary units), both as a function of tip-sample distance $z$, computed over different sites of the sample as indicated by the red and blue dots in the PTCDA structure formula. Experimental images in panels (c) and (e) are taken from reference \cite{Kichin2013}). All simulated images except (a) are normalised to obtain maximum contrast. }
\label{fig-02}
\end{figure}

On top of the mechanical AFM model we derive a~simple numerical model for STM simulations, the main objective of which is to understand the variation of the tunnelling current as a function of \PP\ relaxation. The model is based on Landauer theory \cite{Landauer_IJRD57}. We start from the Landauer formula for the conductance $dI/dV(\epsilon) = \frac{2e}{h} \Gamma_T(\epsilon) {G_P}^\dag(\epsilon) \Gamma_S(\epsilon) G_P(\epsilon)$, where $G_P(\epsilon)$ is the Green's function of the \PP\ at energy $\epsilon$ and $\Gamma_{T,S}(\epsilon) = 2\Im \Sigma_{T,S}(\epsilon)$, where the $\Sigma_{T,S}(\epsilon)={t_{T,S}}^\dag G_{T,S} t_{T,S} $ represent the self-energies of tip (T) and sample (S), respectively;  the $t_{T,S}$ quantify the hopping between the \PP\ and tip or sample, respectively.  To simplify our model we adopt several approximations:  (i) We are interested just in the conductance at zero bias voltage, so we set $\epsilon=\epsilon_F$, where $\epsilon_F$ is the Fermi energy. (ii) We neglect the real part 
of the \PP\ 's Green's function, i.e.~$\Re G_P(\epsilon_F) \approx 0$, and we express the local density of states (LDOS) of the \PP\ as $\rho_P(\epsilon_F) = \frac{1}{\pi} \Im G_P(\epsilon_F)$. In a similar way,  we can also rewrite $\Gamma_{T,S}(\epsilon_F) \approx {t_{T,S}}^\dag(\epsilon_F) \rho_{T,S}(\epsilon_F) t_{T,S}(\epsilon_F)$, where $\rho_{T,S}(\epsilon_F) = \frac{1}{\pi} \Im G_{T,S}(\epsilon_F)$ denote the LDOS of tip and sample, respectively.  (iii) We consider all tunneling channels between the \PP\ and individual atoms of the sample as independent. Thus we can write $\Gamma_S = \sum\limits_{i} \Gamma_i$, where $\Gamma_i = {t_i}^\dag \rho_i(\epsilon_F)  t_i$ is the electronic coupling of the $i$-th atom of the sample to the \PP\ (see \reffig{fig-01}).  (iv) We assume that the sample LDOS $\rho_{T,S}(\epsilon)$ is spread homogeneously over all carbon and oxygen atoms of the molecules which make up the sample, i.e.~the LDOS $\rho_i(\epsilon_F)$ of all atoms is the same ($\rho^{C,O}_i = \mathrm{
const.}$); neither hydrogen atoms of the molecules nor atoms of the metallic substrate are considered.  (v) Finally we assume all LDOS ($\rho_P(\epsilon_F)$, $\rho_T(\epsilon_F)$, $\rho_i(\epsilon_F)$) to remain constant during the scanning process. Therefore they are just multiplicative constants which do not affect the tunneling current variation along the tip trajectory. Consequently, the conductance is only a function of the positions of the tip base atom $\vec{r}_T$, surface atoms $\vec{R}_S$ ($\vec{r}_i$ for individual atoms) and the \PP\ $\vec{r}_P$:

\begin{eqnarray}
 dI/dV(\vec{r}_P,\vec{r}_T,\vec{R}_S) &\propto T_T(\vec{r}_P,\vec{r}_T) T_S(\vec{r}_P,\vec{R}_S) \nonumber \\
 &= T_T(\vec{r}_P,\vec{r}_T) \sum\limits_{i} T_i(\vec{r}_P,\vec{r}_i).
\end{eqnarray}

In other words, we can describe the conductance through the \PP\  junction via two terms: (a) tunnelling from the tip to the \PP\ ($T_T \approx {t_T}^\dag t_T $) and (b) subsequent tunnelling from the \PP\ to the sample ($T_S \approx \sum\limits_{i} T_i = \sum\limits_{i} {t_i}^\dag t_i $) (see \reffig{fig-01}). The tunnelling process $T_i$ between a~given sample atom and the \PP\ can be expressed as an~exponential function $ T_i \propto \exp{(-\beta_S | \vec{r}_P - \vec{r}_i | )}$, where $\beta_S$ represents the characteristic decay length of the tunnelling process between the \PP\ and sample atom $i$. Similarly, we can define $T_T \propto \exp{(-\beta_T | \vec{r}_P - \vec{r}_T | )}$.  An angular momentum dependence of the hopping can also be included (cf.~Supplementary Methods and Supplementary Figure 5). Because we are only interested in the variation of the atomic STM contrast due to the \PP's relaxation, we consider for simplicity both characteristic decay lengths $\beta$ independent of the tip-sample 
distance and equal $\beta_S=\beta_T=\unit{1}{\angstrom}^{-1}$.  We can plot maps of the tunnelling processes $T_S$ and $T_T$ separately to analyse the effects of tip-probe and sample-probe relaxation qualitatively and irrespective of the sizes of the two $\beta$. In reality, $\beta_S$ and $\beta_T$ may differ from each other and according to equation (1) their relative sizes will influence the relative impact of $T_S$ and $T_T$ on the conductance image. 

Clearly, our numerical model omits many processes that happen during tip-sample interaction (e.g.~variations of the LDOS or the tunnelling barrier, multiple scattering effects etc.~\cite{JelinekJPCM12}). Moreover, at close tip-sample proximity additional mechanical degrees of freedom of the junction, such as relaxations inside the tip, within the molecular layer and the surface, will eventually become important. Nevertheless, as will be demonstrated here, this simple model accounts for most of the observed contrast features, which proves the crucial importance of \PP\ relaxation also for the high resolution STM contrast.


\section{Results and Discussion}
\subsection{ High resolution AFM contrast: Inversion and sharpening}

We start the discussion by showing the equilibrium position of the \PP\ when a model tip decorated by CO (`CO-tip') is scanned over PTCDA molecules in the herringbone monolayer. \reffig{fig-02}a clearly shows that at closer distances the functionalized tip experiences sidewise distortions. The observed lateral deformations are induced by the Pauli repulsion that acts at short distances between the \PP\ and the atoms of the PTCDA layer: The \PP\ tends to relax away from the areas where the Pauli repulsion is strong. According to \reffig{fig-02}a, the Pauli repulsion potential over the PTCDA forms `basins' which become clearly visible as the tip approaches the sample.

Let us now inspect the effect that the tip deformation has on the frequency shift $\Delta f$. The maps of $\Delta f$ in \reffig{fig-02}b, calculated at the same tip-sample distances as \reffig{fig-02}a, clearly show the \textit{inversion} of the $\Delta f$ contrast when the tip approaches the sample. Since the observed evolution of the simulated $\Delta f$ images closely matches the experiment (see \reffig{fig-02}c), we can identify the mechanism that drives the experimentally observed inversion of the $\Delta f$ image contrast by analysing the calculated $F_{z}(z)$ curves shown in \reffig{fig-02}f. 

The two $F_{z}(z)$ curves calculated for the tip approach over the carbon atom (red) and over the centre of the aromatic ring (blue) clearly show that initially the repulsion over the carbon atom increases faster upon tip approach. The situation changes at the distance $z\approx\unit{7.4}{\angstrom}$ (\reffig{fig-02}f) when the \PP\ starts to move laterally in order to minimize the effect of the increasing repulsive force. Finally, the repulsion over the ring centre becomes stronger, because there the tip is located in the middle of a potential `basin' that hinders lateral relaxations of the \PP\ and thus prevents the relief of the repulsive force. 

Comparing the simulated force curves in \reffig{fig-02}f with corresponding experimental ones reported in reference \cite{Kichin2013}, we find that both exhibit a very similar behaviour. Small differences between the published experimental $F_z(z)$ curves and the simulated ones shown here can be explained by two facts. First, in our present simulations we do not take into account the attractive interaction between the metal atoms of the tip and the sample. That results in the absence of the attractive force that appears in the experiment after the \PP\ has relaxed laterally out of the junction. Secondly, in the simulation we position the tip precisely over the carbon atom and do not take into the account the finite amplitude of the qPlus oscillation. This produces a sharp kink in $F_z(z)$ at the moment when the \PP\ starts relaxing laterally at the distance $z\approx\unit{7.4}{\angstrom}$ (see \reffig{fig-02}f). Despite these small and well-understood discrepancies between the experimental and simulated 
force curves the overall good agreement between both allows us to conclude that the $\Delta f$ inversion observed both in experiments and simulations occurs due to the decrease of the repulsive force produced by the lateral relaxation of the \PP\ in the junction.

Having shown that our model captures the $\Delta f$ inversion correctly, we note that the inversion effect develops together with a~considerable {\it sharpening} of various features in the $\Delta f$ images (see the middle and the right panels of \reffig{fig-02}b and Supplementary Video 1). Hence, the evident sharpening of the experimental $\Delta f$ contrast at closer tip-sample distances can also be attributed to the increasingly pronounced lateral relaxations of the probe particle. More interestingly, sharp lines also become visible in the intermolecular regions \reffig{fig-02}b where no covalent bonds exist. 

\begin{figure}
\centering
\includegraphics[width=8.5cm]{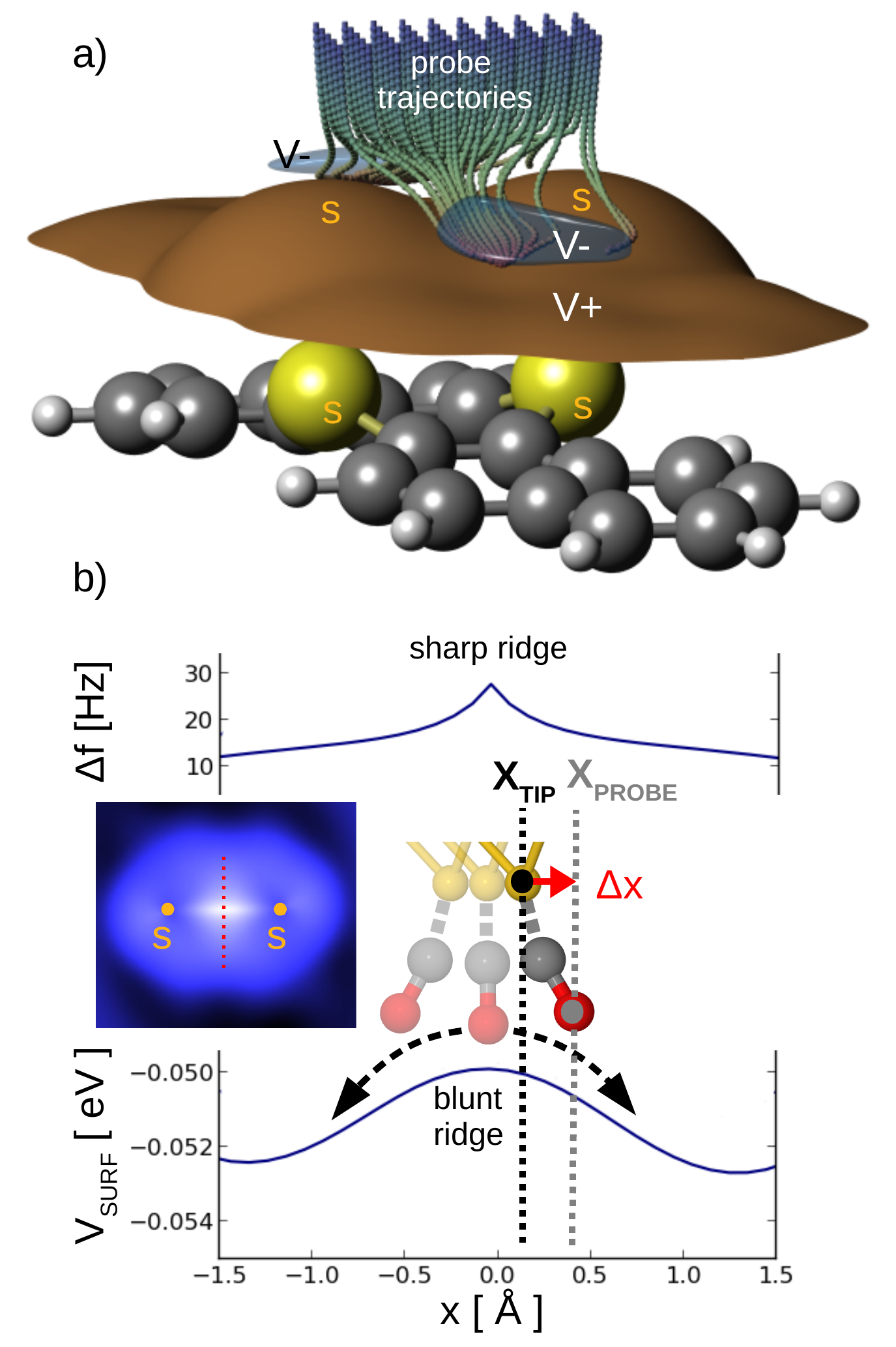}
\caption{\textbf{Origin of sharp lines in AFM images.} (a) DBTH molecule (sulfur atoms in yellow, carbon atoms in dark grey, hydrogen atoms in light grey), with repulsive potential felt by the probe particle in brown ($V+$) and attractive basins ($V-$) in blue. Probe particle trajectories upon tip approach are also shown. Sulfur-derived hillocks in the repulsive potential are labelled `S'. Between them, a repulsive saddle is formed. (b) Surface potential $V^{\mathrm{Surf}}$ (bottom) and AFM frequency shift $\Delta f$ (top) along the central cross section of the repulsive saddle. In the centre the relaxation $\Delta x$ of the probe particle towards the position $x_\mathrm{probe}$ is shown schematically for a tip position $x_\mathrm{tip}$ close to the ridge of the saddle. The mapping of the force at $x_\mathrm{probe}$ to the macroscopic tip position $x_\mathrm{tip}$ explains the sharpening of the $\Delta f$ curve (`sharp ridge') even for a smooth saddle in $V^{\mathrm{Surf}}$ (`blunt ridge'). The inset shows 
a~simulated AFM image of DBTH in the region of the two sulphur atom, clearly exhibiting the sharp line between the two sulphur atoms. }
\label{fig-sharp}
\end{figure}

The origin of sharp lines in AFM images between atoms is schematically illustrated in \reffig{fig-sharp} for the example of the DBTH molecule from reference \cite{Pavlicek2013}, for which such a sharp line is observed between two sulfur atoms which are \textit{not} covalently bonded (see reference \cite{Pavlicek2013} for the experimental image and \reffig{fig-sharp}b (inset) for the simulation). In \reffig{fig-sharp}a the simulated repulsive potential and the attractive basins felt by the \PP\ above the DBTH molecule in the vicinity of the two non-bonded sulphur atoms are shown in a three-dimensional plot, together with the trajectories of the \PP\ as the tip approaches the sample. One clearly observes a repulsive saddle between the two sulphur-derived hillocks (a cross section through this \textit{smooth} saddle is displayed in the bottom diagram of \reffig{fig-sharp}b). The trajectories in \reffig{fig-sharp}a reveal that the \PP\ relaxes away from the saddle ridge. This means that, e.g., for the tip 
position $x_\mathrm{tip}$ in \reffig{fig-sharp}b the \PP\ is subject to the repulsive force from the sample at $x_\mathrm{probe}$. The mapping of forces at $x_\mathrm{probe}$ to the macroscopic tip coordinate $x_\mathrm{tip}$ introduces the sharp ridge in the frequency shift signal $\Delta f$ that is shown in the upper part of \reffig{fig-sharp}b, although the surface potential $V_{\mathrm{Surf}}$ has a smooth saddle.

The mechanism illustrated in \reffig{fig-sharp} is operational and lead to sharp lines in the images whenever a repulsive saddle occurs in the potential felt by the probe particle. The origin of this saddle can either be the presence of `real' electron density or the close proximity of atoms.  In the case of covalent bonds, the electron density also show up as a smooth feature in AFM images recorded at large tip distances at which the probe particle does not show appreciable relaxation (first column of \reffig{fig-02}a-b). In the other case, the saddle arises merely from the convolution of the electron densities of the \PP\ and the neighbouring sample atoms at close tip-sample distances and will disappear for larger distances.

Note that at very close distances the molecular contrast in the experimental images may sometimes become significantly asymmetric, see \reffig{fig-02}c. We attribute this observation to an~asymmetry of some of the CO-tips. To confirm this hypothesis, we have repeated the simulations using a~tilted probe, where the equilibrium position of the probe particle is displaced by 1 \AA \   along $y$-axis away from the lateral position of tip base. The resulting images show asymmetric contrast on benzene rings and in intermolecular features in good agreement with the experimental findings, see Supplementary Figure 4 and Video 1,2. 

\subsection{ High resolution STM contrast on and between molecules}
We now analyze the high resolution STM contrast (i.e.~ STHM- and similar atomic probe contrasts) \cite{Temirov2008, Weiss2010a, Kichin2011, Kichin2013}. This contrast has two aspects. On molecules their geometric structure becomes visible, similar to high resolution AFM \cite{Gross_Science09}, while between molecules very pronounced intermolecular features, such as sharp lines between oxygen and hydrogen atoms or sharp-edged trapezoids between the perylene and anhydride sides of two PTCDA molecules, appear \cite{Kichin2011, Kichin2013}. The first aspect regularly manifests itself very clearly for H$_{2}$-, D$_{2}$- and CO-functionalized tips, whereas the latter is more optimally pronounced with H$_{2}$-, D$_{2}$- and Xe-tips \cite{Kichin2013}.  As we show now, both aspects of high resolution STM contrast are closely linked to the same probe particle relaxation that also governs high resolution AFM images.

First, we focus on the contrast on the molecules, employing exemplary images of PTCDA/Au(111) displayed in \reffig{fig-02}e that were observed in experiments with a CO-functionalized tip. Using the generic transport model described above, we find that the spatial variation of the tunnelling $T_S$ between the \PP\ and the sample (\reffig{fig-02}d) exhibits all essential features of the experimental images in \reffig{fig-02}e. In particular, the overall shape of the molecules and at close distances the appearance of sharp contours between the carbon rings are both reproduced very well. Thus, we are led to the conclusion that in this case of a CO-tip the experimentally observed high-resolution STM images are mainly determined by the $T_S$ tunnelling process.


\begin{figure}
\centering
\includegraphics[width=12.5cm]{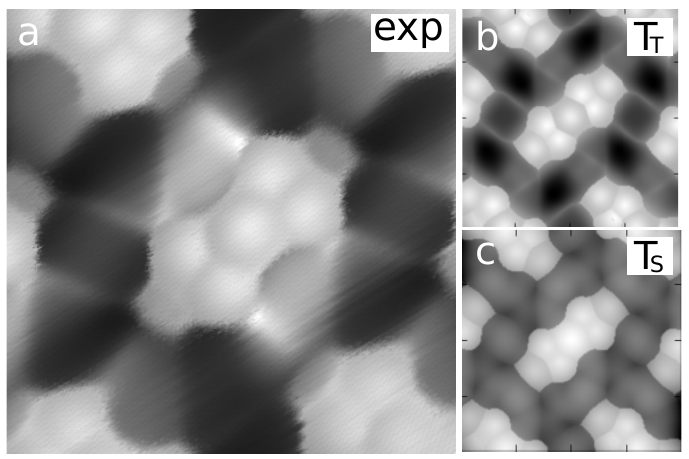}
\caption{ \textbf{Intermolecular contrast in high resolution STM images.}  (a) Experimental constant height image recorded with Xe-tip over PTCDA/Ag(111). Imaging parameters: area: 18$\times$18 nm$^2$, $V=-4$ mV. Prior to the imaging the tip was stabilised at $I=0.1$ nA, $V=-350 $mV, then the bias was changed to $V=-4$mV and the tip was moved by 4 \AA\ closer to the surface. (b) Simulated STM image with Xe \PP\  ($T_T$ tunnelling channel). (c) Simulated STM image with Xe \PP\  ($T_S$ tunnelling channel). Panels (b) and (c) display approximately the same area as panel (a). }
\label{fig-03}
\end{figure}

Turning next to the remarkable STM contrast in the regions between the molecules, we choose the example of a Xe-tip (\reffig{fig-03}). This time we find excellent agreement between experiment (\reffig{fig-03}a), carried out on PTCDA/Ag(111), and the simulated image of $T_T$ tunnelling (\reffig{fig-03}b).  Also, the intramolecular contrast (bright aromatic rings with sharp C-C bonds appearing dark) is well reproduced in the $T_T$ channel, although on the molecules the difference between the $T_T$ and $T_S$ images in \reffig{fig-03}b and \reffig{fig-03}c is not so large. 

Apparently, different experimental situations generate tunnelling contrasts of either $T_S$ or $T_T$ type. The obvious questions is why? The variation of the tunnelling current in each channel depends exponentially  on the distance between the \PP\ and the tip base ($T_T$) or the surface atoms ($T_S$). In the case of a CO-tip, the presence of the stiff covalent bond between CO and the tip base implies only minor changes in the tip-CO distance as the tip is scanned across the sample. Consequently, the $T_T$ channel does not contribute to the STM contrast significantly and the $T_S$ channel prevails. 

The situation is different in the case of a Xe-tip. The weak interaction (i.e., less stiff bond) between the Xe atom and the tip base leads to more contrast in the $T_T$ channel for Xe than for CO. At the same time, the contrast in the $T_S$ channel will be reduced for Xe relative to CO, for two reasons: Firstly, while CO is electronically more strongly coupled to the tip (i.e., low $\beta$) than to the sample, the electronic couplings of the Xe atom to the tip and the sample are expected to be rather comparable; this reduces the relative importance of the coupling to the sample. Secondly and even more importantly, the large atomic radius of Xe smears out the variation of the surface potential, and thus the contrast in $T_S$, effectively. In conclusion, the mechanically and electronically more weakly coupled part of the junction tends to determine the high resolution STM image.

\subsection{ Can hydrogen bonds be imaged? }

The striking AFM/STM contrast between molecules, including the sharp lines observed there, appear suggestive of intermolecular bonds \cite{Weiss2010, Kichin2011, Kichin2013, Zhang2013, Sweetman2014}. We therefore proceed with addressing in the framework of our mechanical model the imaging of \textit{hydrogen bonds} with AFM that was attributed to the enhanced electron density between oxygen and hydrogen atoms of neighbouring molecules \cite{Zhang2013}. 

\begin{figure}
\centering
\includegraphics{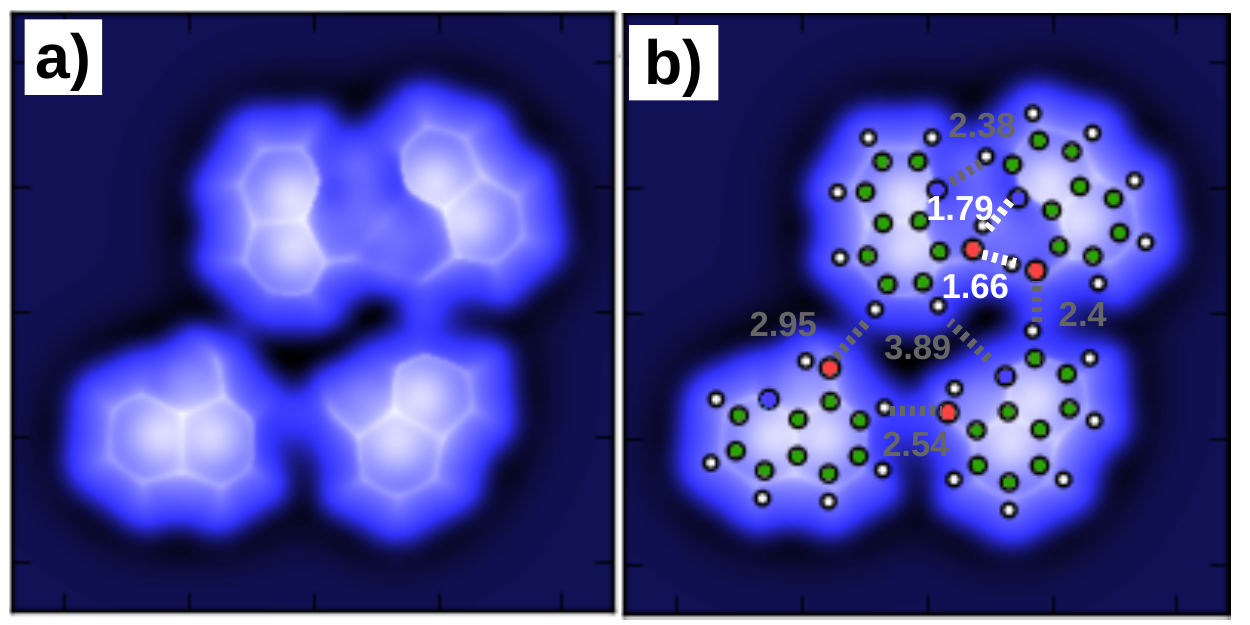}
\caption{ \textbf{Sharp intermolecular contrast and hydrogen bonds.} (a) A simulated AFM image for a 8-hydroxyquinoline tetramer with CO probe particle (tip-sample distance 7.4 \AA). The sharp lines in intermolecular regions  agree very well with the contrast reported in Fig.2d of reference \cite{Zhang2013}. (b) Same image as in panel (a), but with a schematic overlay of the molecular structure, with atoms discriminated by colours (white: hydrogen, green: carbon, blue: nitrogen, red: oxygen), and hydrogen bonds (short and strong hydrogen bonds: white, weaker hydrogen bonds: grey). Numbers indicate bond lengths in \AA. The image colour scale is rescaled by maximum and minimum values of $\Delta f$ to provide best contrast. }
\label{figHQ8}
\end{figure}

Our previous discussion of the results displayed in Figs.~\ref{fig-02}, \ref{fig-sharp}, \ref{fig-03}  has shown quite generally that the intermolecular contrast is very closely related to the lateral relaxation of the probe particle. We are therefore led to the suggestion that also the `hydrogen bonds' of reference ~\cite{Zhang2013} may in fact be due to this effect. To test this conjecture that the observation of {\it apparent bonds} is in general mainly driven by the relaxation of the probe particle, we have performed AFM simulations for the 8-hydroxyquinoline tetramer (\reffig{figHQ8}) molecule that was investigated by Zhang et al.~\cite{Zhang2013}. The~experimental image depicted in Fig.2B in reference \cite{Zhang2013} should be directly compared to our simulated image at the distance $z$=7.4 \AA\ (\reffig{figHQ8}b). Our simulation resolves sharp intermolecular lines connecting typical donors (-OH groups) and acceptors (N, O atoms) of hydrogen bonds very well (\reffig{figHQ8}). According to the 
mechanism discussed in the context with \reffig{fig-sharp}, these lines are well resolved in our model simply because of the close proximity between donor and acceptor atoms (see white dotted lines on (\reffig{figHQ8}d) with bond length labels). Certainly, our purely mechanical model does not have an increased electron density along those lines. On the basis of this finding we suggest that also in the experiments of Zhang et al.~\cite{Zhang2013} probe particle relaxation may in fact be the origin of the observed features which the authors identify with hydrogen bonds. 

The longer hydrogen bonds between CH groups and O, N atoms (grey dotted lines \reffig{figHQ8}d), which are less pronounced in Fig.2 B of reference \cite{Zhang2013}, become visible in our simulations only at even closer tip-sample proximity (cf.~Supplementary Figure 6). This as well as some other minor discrepancies with experimental image (namely the different distortion of aromatic rings), can be explained naturally by two reasons: (a) The positions of the atoms in the molecules in our input geometry are probably not exactly the same as in experiment. (b) In the experiments, the molecules can move slightly on the surface in both lateral and vertical direction under forces exerted by the tip. These degrees of freedom are not included in our simulation.  

\begin{figure} [!ht]
\centering
\includegraphics[scale=0.6]{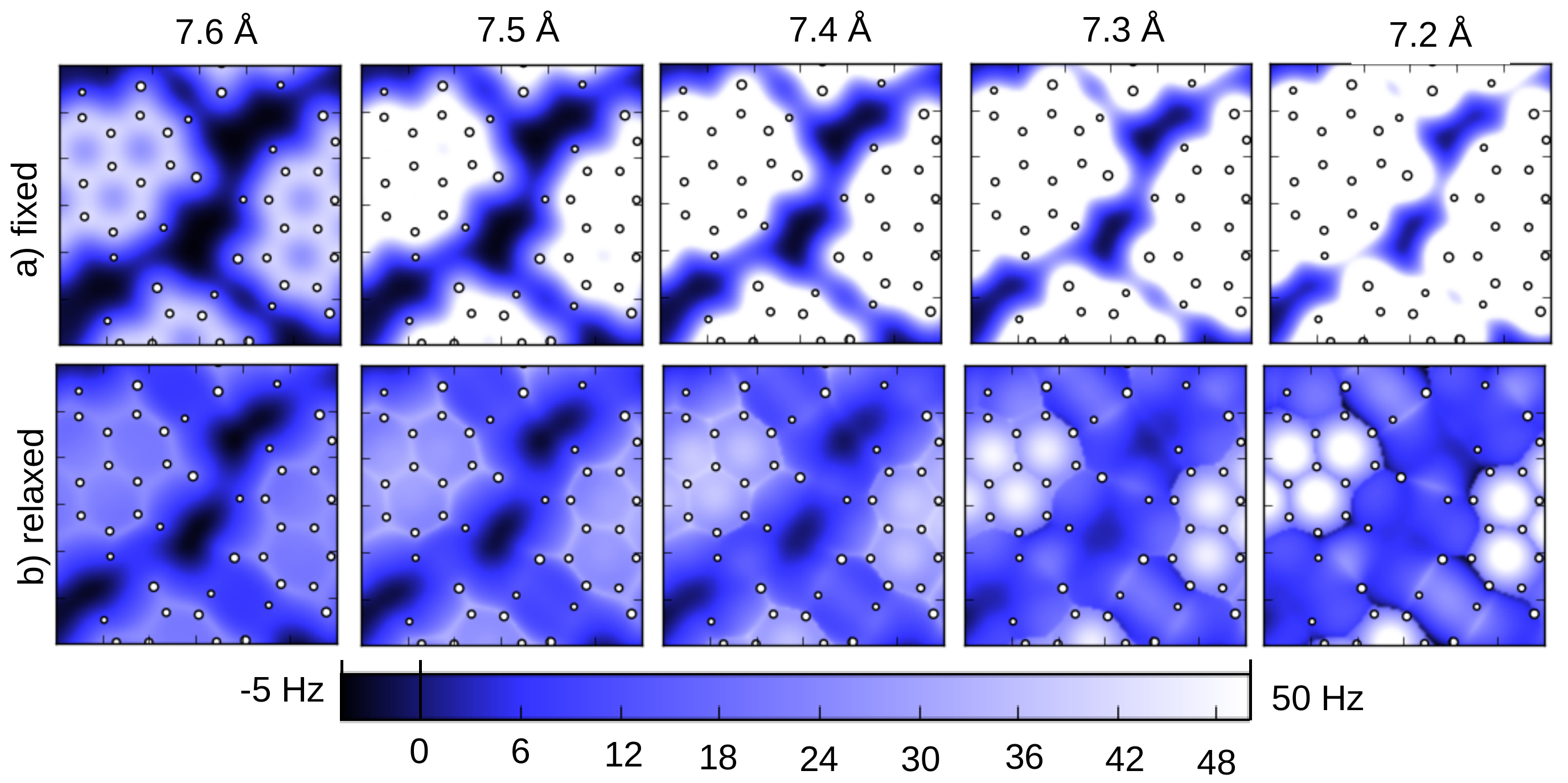}
\caption{ Our simulated AFM images of monolayer of naphthalene tetracarboxylic diimide (NTCDI) molecules which was experimentally studied by \cite{Sweetman2014}. Figs a) show the simulation without consideration of \PP\ relaxation. Figs b) shows simulation with relaxation of \PP  which should be directly compared with Figure 1b,c of \cite{Sweetman2014}; Both images are in the same color scale of frequency shift corresponding to cantilever stiffness 1800.0 N/m and basic frequency 30.3 kHz.}
\label{figNTCDI2}
\end{figure}

Recently an alternative explanation of the origin of the hydrogen bonds was proposed by \cite{Sweetman2014}. They attribute  the imaging mechanism of the hydrogen bonds to a~change of the electron density upon tip approach without consideration of a~tip relaxation. We would like to illustrate, that even though mechanism discussed in \cite{Sweetman2014} may be present, the probe relaxation is the driving mechanism, which makes the intermolecular bonds visible in the AFM experiment. To do so we compare experimental AFM images acquired over naphthalene tetracarboxylic diimide (NTCDI) \cite{Sweetman2014} to images calculated using our model with ( \reffig{figNTCDI2}a ) and without ( \reffig{figNTCDI2}b ) \PP\ relaxation. Although in both cases we can observe the increased repulsion in region between oxygen and hydrogen atom, only the results obtained with relaxing \PP\ are similar to the experimental evidence. In simulated images for fixed \PP\ the variation of the repulsion over the bond is very smooth ( 
similarly to what is shown in Figure 4f of \cite{Sweetman2014}) and the magnitude of the repulsive interaction over hydrogen bonds much smaller than over the molecule. Only if we take into account the \PP\ relaxation  we are able to reproduce the sharp contrast visible simultaneously over the molecules and the intermolecular region in the experiment.

In addition, we would like to note that the observation of the enhanced Pauli repulsion over hydrogen bond with rigid tip cannot be directly related to the increased density on intermolecular bonds (i.e. between oxygen and hydrogen atom). Similar effect can be achieved by a~convolution process due to finite size of tip (i.e. \PP). Here the Pauli repulsion can be approximated by an~overlap of the electron densities of tip and sample. Even in the case, when there is no electronic density between two atoms on surface (i.e. oxygen and hydrogen in hydrogen bond) due to zero overlap between their wave functions, the probe particle with radius comparable to the bond length can overlap with both surface atoms simultaneously. This gives arise the enhanced repulsion in between the atoms in AFM imaging,as consequence of the superposition of the repulsions steaming from both surface atoms. Based on this argument, it is hard to discriminate the mechanism proposed in \cite{Sweetman2014} from this convolution effect. 


\section{Conclusions}

In conclusion, we have developed a reliable numerical model which despite its simplicity is able to reproduce high resolution AFM and STM images of molecules, recorded with functionalized tips, very well. The excellent agreement between simulated and experimental images allows us to show that the appearance of sharply resolved structural resolution, observed experimentally both in the AFM and the STM mode, is due to strong lateral relaxations of the \PP\ attached to the metallic tip apex. At close tip-sample distances these relaxations follow the potential energy basins produced by the Pauli repulsion. Therefore, sharp features appearing in the images always coincide with the borders of neighbouring basins, i.e.~the narrow areas where the magnitude and the direction of the lateral relaxation of the \PP\ changes strongly upon small variations of the position of the tip relative to the sample. Since the lateral and the vertical relaxations of the \PP\ are closely coupled, in the area between the neighbouring 
basins the vertical position of the \PP\ also becomes very sensitive to the precise position of the tip, thus producing the sharp image features in the AFM images. 

Furthermore, we have also demonstrated that our mechanical model, if combined with a~generic model of tunneling through the \PP\ junction based on the simplified Landauer formalism, successfully explains the features of high resolution atomic contrast of STHM, too. Regarding the STM contrast, we note a few salient points: 
(1) Any extension of the present tunneling model that includes more realistic charge transport effects will not change its essential feature: Namely, that the observed STM contrast is directly related to the abrupt relaxation of the \PP\ when the junction crosses boundaries between neighbouring basins of the repulsive potential. 
(2) Our model confirms the concept of the probe particle acting as a combined sensor and transducer \cite{Weiss2010a, Kichin2011, Kichin2013}. This concept relies on the presence of at least one internal degree of freedom in the tunneling junction which can sense a certain physical quantity and transduce this signal into another physical quantity. In the present case, our mechanical model shows how the probe particle senses repulsive forces and by its response to them (mainly lateral relaxation) couples this signal into the tunneling conductance of the junction. (3) The  high resolution AFM and STM imaging mechanism discussed in this communication can be also applied to point contact microscopy  \cite{Stroscio2004,Zhang2011,Schull2011}, as in all the cases sharp features originate from relaxation of central part of the junction. However, in the present case there is \textit{no hysteresis} in the position of the probe particle as the tip is scanned across the surface. As a consequence, there is also no 
dependence on scanning direction or speed, and all images can be reconstructed from force vs. distance or conductance vs. distance curves measured `statically' (as far as the lateral position is concerned) on a grid above the surface.


Finally, we have demonstrated that sharp image features recorded with functionalized STM/AFM tips, and in particular the resolution obtained in the areas between molecules, do not follow an increased electron density corresponding to any kind of interatomic or intermolecular bonds, but they trace the sharp boundaries between basins of the short-range repulsive potential produced by atoms that reside close to each other.


\section{ acknowledgements }
P.H. and P.J. acknowledge the support by GA\v{C}R, grant no.\ 14-16963J and CAS M100101207. R.T. thanks the Helmholtz Gemeinschaft for financial support in the framework of a Young Investigator Research Group. We thank P. Moriarty (University of Nottingham) and J. Repp (University of Regensburg) for fruitful discussions and valuable comments.


\newpage


%

\end{document}